# Schopenhauer on Space, Time, Causality and Matter: A Physical Re-examination

**Shahen Hacyan**

Instituto de Física,

Universidad Nacional Autónoma de México,

A.P. 20-364, México 01000,

Mexico

**Abstract**

According to Schopenhauer, Kant's arguments about the transcendental ideality of space and time can be extended to matter through the concept of causality and the principle of sufficient reason. In this article, I examine to what extent space, time and causality can be considered *a priori* concepts in the light of classical and modern physics. The concepts of matter and field, and their possible *a priori* fundaments, as stated by Schopenhauer, are thus revisited in a modern context.

|

hacyan@fisica.unam.mx

Kant argued that space and time are given *a priori* as forms of intuition (*Anschauungsformen*), and "transcendental aesthetics cannot contain more than these two elements" (B58).[1] Objects affect our senses as phenomena in space and time, and the phenomena are produced by a *thing-in-itself* which is not directly accessible to our senses.

In *The world as will and representation* (WWR), Schopenhauer accepted Kant's transcendental aesthetics, but reduced the categories of his transcendental logics to only one: causality. He argued that causality is another form of perception, though distinct from space and time, and that it is also given *a priori,* prior to all experience. Thus, he stood in opposition to empiricists such as Locke and Hume, who believed that our concept of causality follows from empirical experience, that is, it is given *a posteriori*.

In his famous doctoral thesis, *The fourfold root of the principle of sufficient reason* (4R), Schopenhauer argued that causality is indissolubly associated to the principle of sufficient reason. He identified four roots of this principle, two of them being related to the perception of matter which he reasoned to be "through and through causality".

In previous works (Hacyan 2004, 2006), I pointed out that the concepts of space and time, as known in the macroscopic world, cannot be applied to the atomic world. The main argument was that the well-established existence of non-local effects in the quantum world is perfectly consistent with Kant´s thesis that space and time are forms of intuition, not to be found in the thing-in-itself producing a phenomenon, but in its intuition.

The aim of the present article is to examine if the concept of causality is given *a priori*, as claimed by Schopenhauer, and to analyze its relationship with the concepts of matter and mass in the light of classical and modern physics. In particular, I will consider the concept of field in modern physics, which has replaced the vague notion of substance as the substratum of matter in classical philosophy.

1. **Causality and sufficient reason**

---

[1] All references to Kant are to the *Critique of Pure Reason.*

According to empiricism, we learn that a certain effect has such and such a cause only by experience. Schopenhauer, on the contrary, argued that causality is a form of understanding given *a priori*: the fact that we always associate a cause to an effect is an innate process.

Consider, for instance, the fall of heavy bodies due to the action of gravity. If it were a purely empirical fact, it could occasionally happen that a body remained in levitation; that would be quite improbable but not fully impossible. After all, in another example, I may be buying a lottery ticket every week during many years without winning the big prize and that would be an empirical fact; but if I win someday, no law of nature would be violated.

Of course, we are certain that bodies fall by the action of gravity. That is a law of nature. If, by any chance, it happened that a body did not fall, we will not be satisfied with the explanation that that is very improbable but not utterly impossible. Quite on the contrary, we would look for a reason for such a phenomenon; for instance, that the said body is lighter than air, that it is hold by invisible strings, that it is levitated by magnetic fields, etc. Actually, that is what science is about: given a new and unexpected phenomenon, one *always* looks for its cause.

Consider a more realistic example: when Kamerlingh Onnes discovered that the electrical resistivity of mercury drops exactly to zero below a certain temperature, physicists spent several decades looking for an explanation of this entirely unexpected phenomenon; they finally found its (sufficient) reason in the realm of quantum mechanics. Another example is the actual situation in cosmology: the observational data fit quite well with the general relativistic model of the Universe, but only assuming the presence of an invisible "dark matter", together with an even more mysterious cosmic acceleration. At present, most cosmologists believe that the model is correct and look for a direct evidence of dark matter. Other researchers have pointed out that gravitation may not behave as believed at very large scales. In either case, a sufficient reason is sought, whether in the form of a new kind of matter or a departure from the currently accepted theory of gravity. In the given examples, causality is closely related to the principle of sufficient reason: there is a reason for every phenomenon.

In his doctoral thesis (4R), Schopenhauer identified as the first root of this principle the causal association of a sense stimuli to the phenomenon producing it. For instance, the physico-chemical stimuli produced by light on the retina and the transmission of this information to the brain through the optical nerve, requires a processing of this crude information by a certain form of understanding in order to yield an image of reality. In the modern language of computers, we could say that the brain has a *software* that processes the information and produces the images of the world. Anachronism apart, Schopenhauer's point is that such a software is innate! We are born with it in our brain and learn to use it through experience.

In Schopenhauer's times, the perception of nature was basically through direct sense perceptions. It is mainly in the nineteenth century that scientist started to study the world with the massive support of sophisticated apparatuses designed to extricate its secrets to nature. Nowadays, physicists who study, say, subatomic particles do not see electrons, protons, Higgs bosons, etc. in a direct way: all they can do is to analyze a huge quantity of data collected by appropriately designed detectors and deduce, through a theoretical model, the processes that took place. Such an analysis must be based on some previously established theoretical model; if the model is wrong, contradictions appear between the collected data and their interpretations; otherwise, only the self-consistency of the model is proved (in the case of the Standard Model of elementary particles, no inconsistency has been found so far).

A century ago, Duhem (1914) discussed this important point in detail:

> "An experiment in Physics is not simply the observation of a phenomenon; it is, in addition, the theoretical interpretation of this phenomenon"

Indeed, a layperson visiting a physicist's laboratory would only see a set of apparatuses and monitors displaying graphs and numbers, but he/she will not guess what it is all about.

Summing up, there are various steps in the process of understanding. In Kantian terms, there is a thing-in-itself that produces a phenomenon. This phenomenon is perceived either a direct stimulus in our sense organs which is further processed by the brain, or produces some data in an apparatus that are in turn processed according to a pre-established theoretical model. In the former case, the relation between cause and effect is provided by an internal innate "software" in the brain; in the latter case, use is made of an external "software" (including, nowadays, a real software in an artificial brain). In modern science, experimental results and their interpretations are necessarily mediated by a theory providing the causal relations between crude data and the theoretical description of a physical phenomenon. If some unexpected result or inconsistency appears, it must have a sufficient reason: it may be something yet unknown within a known theory (such as dark matter) or the need to modify the known theory (such as a modification of general relativity).

## 2. Matter and causality

Matter is quantified as mass, and mass can be determined empirically, either directly as weight or indirectly by the motion of a body. Accordingly, Kant did not consider mass as a pure form of intuition given *a priori*. On this point, he is quite explicit in stating that "the possibility of the synthesis of the predicate 'weight' with the concept of 'body'… rests upon experience" (B12). Schopenhauer, however, in the third root of the principle of sufficient reason declared matter to be the "perceptibility of time and space, on the one hand, and causality that has become objective, on the other" (4R, §35), and furthermore:

> "The eternity of matter follows from the fact that the law of causality refers only to the *states* of bodies… it is by no means related to the existence of *that which bears* these states, and has been given the name of *substance… Substance is permanent…*" (4R, § 20)

He further elaborated on this point in WWR, where, in particular, he postulated the concept of substance as the substratum of matter:

> Pure matter… alone constitutes the true and admissible content of the conception of *substance*, is causality itself, thought objectively, consequently as in space, and therefore filling it. Accordingly, the whole being of matter consists in *acting*. Only thus does it occupy space and last in time. It is through and through pure causality. (WWR Chap 24)

Hence the different manifestations of matter are *accidents* of the substance, or more precisely "a particular mode of action… *in concreto.*" (4R, § 21). On this point at least, Schopenhauer follows Locke, who divided qualities into primary and secondary, the latter being mutable.

Now, causality manifests itself through matter. According to Schopenhauer, the third root of the principle of sufficient reason refers to matter as *pure intuition.* "Matter [is] the perceptibility of time and space, on the one hand, and causality that has become objective, on the other" (4R, § 35). This point is further elaborated in WWR, where Schopenhauer treats matter as if it were a third form of intuition. Accordingly, he included matter in parallel with space and time in his table of *prædicabilia a priori*, which are "all the fundamental truths rooted in our *a priori* knowledge of perception" (WWR Chap. IV of Vol. 2). In his own words:

> Matter… is not *object* but *condition* of experience, just as are space and time. This is why, in the accompanying table of our pure fundamental knowledge *a priori, matter* has been able to take the place of *causality*, and, together with space and time, figures as the third thing which is purely formal, and therefor inherent in our intellect.

At present, some of the *a priori* truths in his table can be questioned as not making much sense in modern science, but in most cases their *a priori* nature is worth examining… even at the risk of falling into anachronisms!

Thus, for instance, the first *a priori* truth states: "there is only one time… -- only one space… --only *one* Matter, and all different materials are different states of matter; as such it is called *Substance"*. Of course, the concept of substance used by ancient philosophers is nowadays completely outdated, but (as examined in the next sections) what Schopenhauer apparently had in mind could be closer to the modern concept of field in physics. This is consistent with his second *a priori* truth: "different matters are not so through substance but through accidents", provided, of course, we interpret "accident" as a particular manifestation of physical fields.

Then the third *a priori* truth, "time cannot be thought away… —space cannot be thought away… -- The annihilation of matter cannot be conceived, yet the annihilation of all its forms and qualities can" may be given a modern interpretation: matter *and energy* can transform into each other, but the total matter-energy cannot be annihilated (for instance, an electron and a positron produces two gamma rays, particles of pure energy). I will return to this point in section 3 below.

The fourth *a priori* truth is: "Matter exists, *i.e.*, acts in all the dimensions of space and throughout the whole length of time, and thus unites and thereby

fills these two. In this consists the true nature of matter. It is therefore through and through causality."

The fifth *a priori* truth is that time, space and matter are each infinitely divisible. Of course, we now know about quarks, electrons and other elementary particles, but in physical theories these are described as point particles having no structure and therefore admitting no further division;[2] if some structure were discovered in the future, more elementary constituents will be looked for.

The sixth truth is that time, space and matter are homogeneous and form a continuum. This seems to contradict what we know about elementary particles, but the field of modern physics does form a continuum (see Section 4 below).

Finally, let us discuss at some length the eighteenth truth:

> Time is not measurable directly through itself, but only indirectly through motion, which is in space and time simultaneously; thus time is measured by the motion of the sun and of the clock.
>
> Space is measurable directly through itself, and indirectly through motion, which is in time and space simultaneously; thus, for example, an hour's walk, and the distance of the fixed stars expressed as so many light years.
>
> Matter as such (mass) is measurable, *i.e.*, determinable according to its quantity, only indirectly, thus only through the *magnitude of the motion*, which it receives and imparts by being repelled or attracted.

Time was measured in the past with the periodic motion of celestial bodies, and nowadays it is measured with the periodic vibrations of atoms. Space is presently measured in terms of the distance covered by light in a (well defined) unit of time. As for mass, its quantification is more problematic; the previous eighth *a priori* truth states that "by reason of matter we weight", but it has been considerably more difficult to find a standard of mass in terms of purely natural constants.[3]

---

[2] "Superstrings" have some structure, but in an abstract mathematical space.
[3] Hopefully, the Sèvres standard will be soon replaced by an atomic standard based on the value of the Planck constant.

## 3. Space, time and mass

There are several definitions of mass in physics textbooks. It is usually defined as the "quantity of matter", albeit it is never specified how such a quantity can be measured. Moreover, there is a general confusion between mass and weight. We now know that weight, though related to mass, is a manifestation of the gravitational force and vanishes in outer space. Obviously, this was unknown in antiquity, and there was even confusion between size and weight due to the fact that bigger bodies are usually heavier than smaller ones (Jammer 1961).

In the *Principia*, Newton defined mass as the "quantity of matter", which is "the measure of the same [matter], arising from its density and bulk conjointly." It appears, therefore, that for Newton density was a more primary concept than mass. As for Newton's second law, its usual textbook form, "force equals mass times acceleration", is due mainly to Euler. Euler argued that the primary concept should be force and not mass: for how can mass be measured without observing the motion of a body experiencing a certain force of known magnitude? Accordingly, Euler postulated mass as the ratio of force to the acceleration it produces.

However, it is not obvious that force should be a primary concept. Jammer (1961) remarked that, due to the new positivistic attitude: "What once, in Newtonian physics, played a central role was now regarded as an obscure metaphysical notion that has to be banished from science." Is it then possible to measure mass without referring to force (or gravity)? Ernst Mach (1893) conceived an idealized scheme to deduce the mass of two bodies from their motions in a collision, but the method is not at all practical.

Summing up, while the concept of acceleration, which is given in terms of space and time, is intuitively clear, it is not clear that force or mass should be primary concepts. One can either define mass as the ratio between force and acceleration, or define force as the product of mass and acceleration.

We are therefore faced with the fact that mass can be measured only through motion (or equilibrium) in space and time, as anticipated by Schopenhauer in its table of *prædicabilia*.

## 4. Substance and matter

Classical philosophers called "substance" the underlying and permanent element of the world. Kant though that it is necessary to "presuppose its existence throughout all time" (B 228), since "the unity of experience would never be possible if we were willing to allow that new things, that is, new *substances*, could come into existence" (B 229).

Kant described the principle of conservation of substance as an "analogy of experience". Regarding mass, he gave the example of how a philosopher would determine the weight of smoke: "Subtract from the weight of the wood burnt the weight of the ashes which are left over, and you have the weight of the smoke". This, however, as Kant pointed out, is based on the presupposition that "matter (substance) does not vanish, but only suffers an alteration of form" (B 228). This should be known *a priori*, following from a principle of permanence, even though it can be experimentally confirmed or disconfirmed *a posteriori*. If disconfirmed, some sufficient reason for the discrepancy will be looked for without abandoning the general principle.

Now, as for a principle of permanence, physicists would rather refer to the conservation of mass-energy. The fundamental concept of energy, together with its conservation law, appeared in physical theories in the middle of the nineteenth century. It is only then that physicist realized that there must be, besides matter, some conserved quantities in the physical world. From then onwards, the concept of force, which seemed to be so fundamental, was gradually substituted by the more abstract but mathematically well-defined concept of energy. Furthermore, when Einstein proved the equivalence between mass and energy, it became clear that what is conserved is not mass or energy separately, as previously believed, but mass and energy together (for instance, it was realized that an atomic nucleus has slightly less mass than the sum of its constituents, the difference being the binding energy).

Although the conservation of mass-energy is presently well accepted, it is nevertheless an empirical principle of physics, and, as such, Kant would say that it lacks "strict universality and apodictic certainty". An historical curiosity may illustrate this point: When it was found out that energy was apparently not conserved in nuclear beta decays, no less an authority than Niels Bohr proposed the daring hypothesis that the conservation of energy is a statistical principle only and does not apply at the atomic level. The alternative proposed by Wolfgang Pauli was to keep this principle at the expense of postulating the existence of an unknown particle, invisible and intractable, which would be produced in a beta decay and carry the missing energy away. Thus, in a sense, Pauli was following the method of Kant's hypothetical philosopher: instead of smoke, he weighted an even more elusive object, which turned out to be the neutrino. At the time of this discussion, both alternatives seemed equally convincing, but experiments finally confirmed Pauli's hypothesis.

Of course, it would be an anachronism to interpret matter and substance as understood by classical philosophers in terms of modern physical concepts. But the important point I want to stress is that the existence of something acting as the universal substratum of *all* material phenomena must be a belief *a priori*, since it cannot be proved empirically. A conservation law is always wanted, even though it may evolve according to newly discovered facts. And as for the substratum, something equivalent can be found in the concept of field in modern physics.

## 5. Matter and field

A fundamental concept forged by physicists was that of *field*. In this respect, Einstein noted that (Jammer 1954):

> The concept of the material objects was gradually replaced as the fundamental concept of physics by that of the field. Under the influence of the ideas of Faraday and Maxwell the notion developed that the whole of physical reality could perhaps be represented as a field whose components depend on four space-time parameters.

Material objects interact between themselves in different ways and it is the most important purpose of physics to explain their interactions. As mentioned by Einstein, this was achieved in the nineteenth century for electromagnetism with the concept of field. Another force of nature, gravity, can also be described in terms of a field, as in general relativity (the gravitational field can be interpreted as a Riemannian space, but that is another story).

With the discovery of the atomic world and the advent of quantum mechanics, it was possible to describe nuclear processes with a quantum theory of fields. Quantum Field Theory (QFT) proved to be an extremely successful theory of all electromagnetic and nuclear processes. In this theory, the elementary particles that constitute matter are interpreted as vibrations of a field; the energy of these vibrations is quantized, and each quantum of energy is identified with a subatomic particle. The so called Standard Model of elementary particles, which is based on QFT, has been confirmed in all possible ways and provides an accurate descriptions of nature at the atomic and subatomic level… even though its success does not cease to surprise physicists!

Basically, QFT is a mathematical theory that describes various kinds of fields as functions of space and time, which are supposed to form a continuum.[4] Space and time, the two forms of intuition according to Kant, are therefore the primary and fundamental concepts of QFT, and the field itself, as any physical quantity, is a function of variables and parameters with units of length, time and mass (or energy).

## 6. On mathematics

Schopenhauer was not fond of mathematics. Without denying its practical use, he was convinced that mathematics could only yield a quantitative description of the material world, but could never provide an understanding of its causal relations. "Where calculating begins, understanding ends" was his statement on this matter (4R, §21; see also Chap. XIII of WWR). His view may seem to be anachronic nowadays, but it must be recalled that his dislike of mathematics was shared by many intellectuals of his time, who longed for a direct perception of nature and its laws without the intermediary of abstract concepts. Let us recall that even Isaac Newton was criticized in his time for having "only" described the motion of planets, without explaining the real cause of gravity. Goethe, a contemporary much-admired by Schopenhauer, was a strong critic of abstractions in the description of nature; they may be useful, he held, but "it does not occur to the architect to pass off his palaces as mountain sides and forests" (cited by Heisenberg 1990). Even among physicists, the case of Michael Faraday is noteworthy: he performed many crucial experiments that were the basis of a full theory of electromagnetism, but his knowledge of mathematics was quite limited and he deliberately avoided mathematical description in the treaties he authored.

---

[4] There have been some theoretical intents to quantize space and time, buy without clear results.

Nowadays, we are accustomed to the enormous success of mathematics in describing physical phenomena. Nevertheless, Schopenhauer was right in a certain sense; indeed, a mathematical description yields only numbers, but not a real understanding. Quantum mechanics is an excellent example: it is based entirely on abstract concepts, such as wave function, evolution operators, probability amplitudes, collapse, spin, etc., and it has proved to be the most precise description of physical phenomena. Yet, a description in terms of known concepts, such as particle, space and time, has proved to be impossible; any attempt to "explain" quantum mechanics has inevitably led to paradoxes.

Why is mathematics so effective is a great mystery of modern physics. Indeed, its effectiveness is quite unreasonable, as Eugene Wigner (1960) has well pointed out. In conclusion, we could paraphrase Schopenhauer and assert that "where understanding ends, calculating begins…but it may go very far!".

## 7. Conclusions

All equations of physics and all measurements, in any system of units, involve space, time, and mass (or equivalently energy), and, accordingly, the irreducible elements of any system of units in physics are the standards of length, time and mass. The relation between space and time became manifest with the theory of relativity: the connection between these two forms of perceptions is given by the velocity of light (in vacuum), which is a universal constant of nature. A further connection with mass was revealed when Max Planck discovered the quantization of energy. Energy is related to the frequency of oscillation of light through the Planck constant, the fundamental constant of the quantum world. In summary, Planck constant and the velocity of light relate mass, space and time among themselves.

Causality is a fundamental part of physics (not to be confused with determinism). In the presently accepted interpretation of quantum mechanics: there is a wave function that describes all the possible states of a particular physical system, and the intervention of an observer causes its collapse to a wave-function describing a single state. Accordingly, there is a sufficient reason for the "collapse of the wave-function". In general, every effect must have a sufficient reason, to be investigated and discovered. Unexpected results are the most interesting ones, since they open the way to the discovery of new phenomena.

Finally, anachronisms apart, the argument is that we need an *a priori* concept for the substratum of the world. Like the substance of classical philosophers, the field in modern physics is the ultimate underlying substance of the world. It is given in terms of space and time, and generates the mass of the subatomic particles. Thus, it can be interpreted as the sufficient reason for the existence of matter.